**Relative Age Effect in Elite Sports: Methodological Bias or Real Discrimination?**

Nicolas Delorme, Julie Boiché & Michel Raspaud

Abstract

Sport sciences researchers talk about a relative age effect when they observe a biased distribution of elite athletes' birthdates, with an over-representation of those born at the beginning of the competitive year and an under-representation of those born at the end. Using the whole sample of the French male licensed soccer players ($n = 1,831,524$), our study suggests that there could be an important bias in the statistical test of this effect. This bias could in turn lead to falsely conclude to a systemic discrimination in the recruitment of professional players. Our findings question the accuracy of past results concerning the existence of this effect at the elite level.

Keywords: Relative Age Effect, Soccer, Discrimination, Bias.

**Relative Age Effect in Elite Sports: Methodological Bias or Real Discrimination?**

During the last two decades, the Relative Age Effect (RAE) has been a widely studied and commented phenomenon in the sport sciences literature (Musch & Grondin, 2001; Cobley et al., 2009). Among elite athletes, this effect is illustrated by an over-representation of players born during the first two quarters (i.e., three consecutive months' period) of a competitive year, and an under-representation of players born during the last two quarters. This biased distribution would be due to the age categories determined by sport organisations. Those institutions traditionally gather young participants in categories of two consecutive years of birth. Even if such system is settled so as to balance the competition between players, it generates important differences in relative age: two children competing in the same category can have up to 23 months of difference in age if they are not born in the same year, while children who are born in the same year can have up to 11 months of difference.

Consequently, children born early in the competitive year are more easily identified as "talented" or "promising" than their counterparts born later in the year, during the detection sessions organised by sport instances (Helsen, Van Winckel & Williams, 2005). Indeed, their initial advantage in relative age is accompanied by a more advanced physical (Delorme & Raspaud, in press) and cognitive (Bisanz, Morrisson & Dunn, 1995) development. Thanks mostly to their more developed physical attributes (e.g., in height, weight or strength), those children and adolescents benefit from a "biased" vision on their potential, which facilitates their recruitment in high level structures. Once this first step is achieved, they can take advantage of an early exposure to elite practice with highly qualified technicians. This access to top level competition is a key element for their future sport career (Ward & Williams, 2003; Williams, 2000), considering the technical and strategic skills it may bring.[1]

This asymmetrical distribution of elite players' birthdates has been observed in various activities, including baseball, cricket, tennis, football or rugby union (Musch & Grondin, 2001; Cobley et al., 2009). Most of the research nevertheless concerned ice-hockey and soccer. Regarding soccer, a RAE has been reported for the professional championship of numerous countries located in different continents: United-Kingdom (Dudink, 1994), Belgium (Helsen, Starkes & Van Winckel, 1998), Spain (González Aramendi, 2007), France and the Netherlands (Verhulst, 1992), Australia, Brazil, Germany, and Japan (Musch & Hay, 1999). Faced to this consistent set of results in elite sport, Musch and Grondin (2001, 163) conclude their review of the literature stating that "taken together, a growing body of research reviewed in this article suggests that RAEs are a pervasive phenomenon in competitive sport".

Some authors qualify this effect as discriminatory for players born late in the competitive year. In this vein, Edgar and O'Donoghue (2005, 1014) underline the potential gains for a tennis player's career, in terms of money, television coverage, recognition, and celebrity lifestyle, and suggest that "it would be desirable that everyone would have an equal opportunity to become a professional tennis player regardless of season of birth". For those authors, even if this discrimination is inadvertent, it needs to be cautiously examined, given the lucrative nature of certain sports. Other arguments concern the fact that sport should enable every child's blossoming and health (Musch & Grondin, 2001). Indeed, the system seems detrimental for certain children's motivation, which may lead them to dropout and does not contribute to the physical activity habits they shall adopt during adulthood. On a more pragmatic plan, certain authors note that the RAE observed, as an artificial consequence of the youth competition structure, generates a loss in potentially talented players, which in the long run contributes to a decrease of level among professional and national teams (e.g., Pérez Jiménez & Pain, 2008).

Because of those potential economical, psychological and health-related outcomes of RAE , the majority of authors agree about the necessity of carrying actions aiming at reducing this phenomenon or even make it disappear. With this regard, it was proposed to establish among young participants new categorisation systems, either based on biological (e.g., Baxter-Jones,1995) or chronological age (e.g., Boucher & Halliwell, 1991; Hurley, Lior & Tracze, 2001; Hurley, 2009), so as to deal with the negative correlates of the differences in relative age. As sport organisations apparently ignore this phenomenon, some authors even called for a direct intervention of the government (e.g., Hurley, Lior & Tracze, 2001).

In sum, the RAE is thus qualified of discriminatory because it put at disadvantage players born late in the competitive year, by reducing their chances to reach the elite. The accumulation of studies reporting such an effect among high level samples is likely to discourage anyone of doubting of the existence of this systemic discrimination. The purpose of the present work is however to test the empirical reality of this discrimination. Indeed, it looks like an important methodological bias appears in the studies reporting a RAE among elite athletes, which questions the validity of the conclusions presented. In the literature, the presence of RAE is traditionally determined by examining whether or not there is a significant difference between the theoretical expected distribution of players by month (or quarter) and the observed distribution, which is done by performing a chi-square goodness-of-fit test or a Kolmogorov-Smirnov one-sample test. Depending on the studies, four strategies can be found in calculating the expected distribution:

(a) An even distribution of birthdates by month or quarter is posited (e.g., Barnsley, Thompson & Legault, 1992). This choice is frequent when the research concerns an international sample.

(b) An even distribution is posited, controlling for the number of days in a month/quarter (e.g., Edgar & O'Donoghue, 2005). Once again, an international sample is often the justification for this choice.

(c) The birthdates statistics by month, gender and year for the corresponding national population are considered, using weighted mean scores (e.g., Helsen, Starkes & Van Winckel, 1998).

(d) The statistics of a European country are considered, using weighted mean scores, in a study concerning a European sample (e.g., Helsen, Van Winckel & Williams, 2005). This procedure is based on the work by Cowgill (1966) suggesting that the distribution by month/quarter is similar among European countries.

In the absence of specific methodological constraints – for instance, an international sample – in order to obtain results as accurate as possible, it is recommended to use the third procedure to calculate the expected birthdates distribution. Whatever the strategy chosen, all of these four procedures may lead to a biased final interpretation. Indeed, whether the reference considered is an even distribution – procedures (a) and (b) – or national standards – procedures (c) and (d) – there is an implicit postulate according to which the birthdates distribution of the licensed players for a given activity is similar to the corresponding national population taken as reference. Yet, to our knowledge, this postulate has never been tested before, for any sport. It is noteworthy that, except a handful cases, elite athletes are issued from the national population of licensed athletes. As it has been underlined, the differences in relative age are accompanied by significant disparities between players of a same age category, in terms of cognitive and physical development. It thus seems logical that sports where physical attributes represent advantages, such as ice-hockey or soccer, will be less attractive for young people born late in the competitive year, and less physically mature.

As indicated above, the unequal distribution of birthdates among elite players would thus be partly due to a selection system that values physical development and discriminates players born late in the competitive year. Though, in order to definitely conclude to the presence of a kind of discrimination, it is necessary to show that the birthdates' distribution of licensed players by month or quarter is identical to the distribution observed in the corresponding national population. If it is shown that there already exists an unequal distribution among the whole population of licensed players in a given activity, the selection system and the recruitment operated by the different professional channels should not be pointed out as responsible for the unequal distribution among the elite. Conversely, a "self-restriction" process inhibiting certain young and preventing them from even beginning the activity, as well as a quick dropout of players born at the end of the competitive year, could in this case account for this phenomenon at the highest level.

Based on the birthdates of all French male players affiliated to the French Soccer Federation (FSF) for the 2006-2007 season, the purpose of this study was twofold. First, we aimed at examining if the distribution of the birthdates in this sample was identical to the one observed in the French population for the corresponding years of birth. Next, we tested whether using all the licensed players versus the national population as reference to calculate the expected birthdates distribution has an impact on the conclusion drawn regarding the distribution observed among elite players (i.e., French players of the first division championship).

Material and Methods

*Data collection*

For the purpose of the present study, the birthdates of all male French players affiliated to the FSF ($n = 1,831,524$) during the 2006-2007 season were collected from the federation database. Foreign players were excluded from the sample, in order to make sure that all participants were subject to the same cut-off date for age categorisation, and to enable a relevant comparison with the birthdates' distribution observed among the French population. The birthdates of players from the French first division championship ($n = 351$) were collected through the rosters of the Professional Soccer League (PSL). Once again, foreign players were excluded from the sample. Among males, the FSF distinguishes 7 age categories: "less than 7 years", "less than 9 years", "less than 11 years", "less than 13 years", "less than 15 years", "less than 18 years" and "adults".

*Data Analysis*

For each of the 7 FSF age categories, as well as for the sample of players from the PSL, the players' birthdates were classified into 4 quarters. Since the cut-off date used to form age categories has been modified by the FSF (Jullien, Turpin & Carling, 2008), the players born before 1982 were classified from Q1 (August-October) to Q4 (May-July) and the players born in 1982 and after were classified from Q1 (January-March) to Q4 (October-December).

Concerning the players affiliated to the FSF, for each age category, the expected distribution was calculated based on the national birth statistics by month and year for males, using weighted mean scores. Those data were obtained through the National Institute of Economical Statistics and Studies. Regarding professional players, two chi-square goodness-of-fit tests were carried out with the two following procedures for the calculation of expected birthdates distribution:

(a) The national birth statistics by month and year for males, using weighted mean scores.

(b) The statistics by month and year observed for the whole corresponding population of male players affiliated to the FSF, using weighted mean scores.

The tests were conducted with the Statistica 6.1 Software (StatSoft Inc.) with a signification threshold fixed at .05.

Results

Table 1 presents the birthdates' distribution by quarter for each age category identified for male players by the FSF, during the 2006-2007 season.

**** Please insert Table 1 near here ****

A chi-square goodness-of-fit test taking the distribution of licensed players as observed distribution and national values as expected distribution reveal statistical differences for all age categories: less than 7 ($\chi^2$ = 566.75, d.f.= 3, $P$<.0001), less than 9 ($\chi^2$ = 237.90, d.f.= 3, $P$<.0001), less than 11 ($\chi^2$ = 269.97, d.f.= 3, $P$<.0001), less than 13 ($\chi^2$ = 346.07, d.f.=3, p<.0001), less than 15 ($\chi^2$ = 619.08, d.f.= 3, $P$<.0001), less than 18 ($\chi^2$ = 752.99, d.f.= 3, $P$<.0001) and adults ($\chi^2$ = 1721.87, d.f.= 3, $P$<.0001). The results reflect a classical RAE with an over-representation of players born in Q1 and Q2, and an under-representation of players born in Q3 and Q4.

Tables 2 and 3 present the analyses of birthdates distribution by quarter for professional players using as theoretical distribution either the corresponding national population or the whole population of licensed players.

**** Please insert Table 2 near here ****

**** Please insert Table 3 near here ****

If the national population is used as reference to calculate the expected distribution, the results indicate a significant RAE among professional players from the PSL ($\chi^2 = 8.31$, d.f.= 3, $P<.05$). Conversely, if the population of licensed players is used to calculate the expected distribution, one cannot conclude to a biased distribution among first division players ($\chi^2 = 4.69$, d.f.= 3, $P<.20$).

Discussion

Because the method traditionally used to determine the presence of a RAE among elite athletes does not evacuate a bias linked to the choice of national standards in the calculation of the expected distribution for the statistical test, the first aim of this study was to compare the birthdates distribution for the whole population of male players affiliated to the FSF, and in the French population. This comparison allowed testing the postulate according to which both are, as a matter of fact, similar.

This study reveals a systematic significant RAE for all age categories distinguished by the FSF, that is to say, an over-representation of players born in Q1 and Q2 and an under-representation of players born in Q3 and Q4. Those results thus indicate that the "classical" methods used to assess and interpret the presence of a RAE may not always be relevant but instead might introduce bias in the conclusions drawn relatively to this phenomenon among the elite. Indeed, the presence or absence of this effect is most of the time examined by looking at the players' birthdates distribution, comparing them to national standards. Such strategy is based on the implicit premises that the population of licensed players for one activity is similar to the national distribution. However, the present results suggest that, at least in the case of French soccer, there already is a disparity in the players' birthdates distribution, from the less than 7 to the 'adults' category.

This has crucial implications and calls for a methodological change in the calculation of expected distributions in studies investigating the RAE. A study aiming at demonstrating this effect with certainty should use as theoretical distribution the one of the population of licensed players, instead of the national corresponding statistics. Indeed, one could hastily conclude that an asymmetrical distribution of birthdates among elite players results from a RAE, whereas in reality it is only representative of the distribution observed in the population of licensed players. The over-representation of players born at the beginning of the competitive year, and the under-representation of those born at the end, may not be a consequence of the selection system valuing physical development of young players, which may put at advantage children and adolescents born during the first months of the year, but could simply be the mimetic expression of the mass of licensed players.

In this perspective, it would be erroneous to conclude to a discrimination toward players born at the end of the competitive year. The potential impact of this bias is far from being negligible and might even lead to opposite conclusions about this phenomenon. In this vein, the results concerning the birthdates distribution of the 351 French soccer players of the first division championship during the 2006-2007 season varies according to the reference used to calculate the expected distribution for the chi-square goodness-of-fit test. When the French population is used, there is a biased distribution, but with the population of soccer licensed players a similar distribution is observed. In the first case, the researcher will be likely to conclude to a discriminative effect due to the mode of recruitment of the various professional instances, whereas in the second case he/she should conclude to the absence of such effect. Because a great majority of the players that reach the elite level comes from the population of licensed players, and went through all the detection and selection sessions operated by federal organisms, we would recommend using the birthdates distribution of this population, and not national data, in order to calculate the expected distribution. This precaution would permit to avoid a bias likely to drastically distort the results obtained.

## Conclusion

If the present results, in themselves, do not demonstrate the absence of a RAE in elite sport, they nevertheless question the accuracy of the tests performed in past studies investigating this phenomenon and consequently the conclusions drawn. With this regard, if a biased distribution already exists among the whole population of licensed players in one activity, it is normal that, by mimicry, such asymmetry also arises among the elite. Taking the national population as reference, one could be prone to hastily conclude to a discrimination operated by the way professional sport organisations proceed to recruit their athletes.

This study further reveals a significant RAE for all age categories distinguished by the FSF. The fact that a biased distribution was systematically observed for regular players calls for additional work examining what mechanisms lead to such effect. We assume that it results from two major processes: first, a phenomenon of "self-restriction" that prevents children and adolescents born at the end of the competitive year to begin to practice this sport; second, higher rates of dropout among those who begin to play but encounter a temporary physical inferiority, compared to players born early in the year who belong to the same age category. Indeed, if the asymmetrical distribution observed at the higher level is not the result of a biased selection, it looks like the one observed for all licensed players reflect a systemic discrimination for young people born late in the competitive year. Given the multiple benefits of moderate sport participation on social acceptance, psychological self-perceptions and health, such discrimination and its mechanisms deserve to be cautiously examined by future research conducted on RAE.

**Table 1.** Season of birth of French male soccer players (2006-2007).

| Category | Q1 | Q2 | Q3 | Q4 | Total | $\chi^2$ | $P$ |
|---|---|---|---|---|---|---|---|
| Adults | 182,945 | 180,229 | 175,775 | 176,111 | 715,060 | 1721.87 | <.0001 |
| (Δ) | (+7,257) | (+8,388) | (-1,922) | (-13,723) | | | |
| Under 18 | 43,920 | 43,771 | 41,799 | 37,430 | 166,920 | 752.99 | <.0001 |
| (Δ) | (+3,712) | (+956) | (-719) | (-3,949) | | | |
| Under 15 | 38,476 | 37,155 | 36,361 | 32,257 | 144,249 | 619.08 | <.0001 |
| (Δ) | (+3,358) | (+582) | (-857) | (-3,083) | | | |
| Under 13 | 42,432 | 43,535 | 42,586 | 39,185 | 167,738 | 346.07 | <.0001 |
| (Δ) | (+2,518) | (+744) | (-637) | (-2,625) | | | |
| Under 11 | 53,187 | 54,469 | 54,861 | 50,967 | 213,484 | 269.97 | <.0001 |
| (Δ) | (+2,337) | (+754) | (-241) | (-2,850) | | | |
| Under 9 | 55,847 | 57,220 | 57,378 | 52,840 | 223,285 | 237.90 | <.0001 |
| (Δ) | (+2,256) | (+660) | (-180) | (-2,736) | | | |
| Under 7 | 50,965 | 51,749 | 52,298 | 45,776 | 200,788 | 566.75 | <.0001 |
| (Δ) | (+2,742) | (+1,340) | (+247) | (-4,329) | | | |

**Table 2.** Season of birth of first division players (2006-2007).

|  | Q1 | Q2 | Q3 | Q4 | Total | $\chi^2$ | $P$ |
|---|---|---|---|---|---|---|---|
| League 1 | 108 | 89 | 81 | 73 | 351 | 8.31 | <.05 |
| (Δ) | (+23) | (-4) | (-8) | (-11) | | | |

**Note**: Δ is the difference between the observed distribution and the theoretical expected distribution.

**Table 3.** Season of birth of first division players (2006-2007).

|          | Q1    | Q2   | Q3   | Q4   | Total | $\chi^2$ | $P$  |
|----------|-------|------|------|------|-------|----------|------|
| League 1 | 108   | 89   | 81   | 73   | 351   | 4.69     | <.20 |
| (Δ)      | (+16) | (-1) | (-6) | (-9) |       |          |      |

**Note**: Δ is the difference between the observed distribution and the theoretical expected distribution.

Footnotes

[1] For an exhaustive presentation on the mechanisms, factors and moderators of the RAE, see the review of the literature of Musch & Grondin (2001).